\documentclass[11pt, a4wide]{amsart}
\usepackage{graphicx}
\usepackage{amssymb,amsmath}
\usepackage{epsfig}
\usepackage{epsf}
\usepackage{float}
\usepackage{a4wide}
\usepackage{graphpap}
\usepackage{setspace}
\newtheorem{teo}{Theorem}[section]
\newtheorem{defi}[teo]{Definition}
\newtheorem{eje}[teo]{Example}
\newtheorem{lem}[teo]{Lemma}
\newtheorem{pro}[teo]{Proposition}
\newtheorem{coro}[teo]{Corolary}
\newtheorem{obs}[teo]{Remark}
\newtheorem{fact}[teo]{Fact}

\def\tV{{\widetilde V}}

\def\bteo{\begin{teo}}
\def\eteo{\end{teo}}
 \def\bdefi{\begin{defi}}
\def\edefi{\end{defi}}
\def\beje{\begin{eje}}
\def\eeje{\end{eje}}
\def\blem{\begin{lem}}
\def\elem{\end{lem}}
\def\bpro{\begin{pro}}
\def\epro{\end{pro}}
\def\bcoro{\begin{coro}}
\def\ecoro{\end{coro}}
\def\bob{\begin{obs}}
\def\eob{\end{obs}}
\def\bfact{\begin{fact}}
\def\efact{\end{fact}}

\def\P{{\mathbb P}}
\def\R{{\mathbb R}}

\def\square{\ifmmode\sqr\else{$\sqr$}\fi}
\def\sqr{\vcenter{
         \hrule height.1mm
          \hbox{\vrule width.1mm height2.2mm\kern2.18mm\vrule width.1mm}
         \hrule height.1mm}}                  

\def\tV{{\widetilde V}}

\def\R{{\mathbb R}}  
\def\P{{\mathbb P}}  
\let\cal=\mathcal

\def\BB{{\cal B}}
\def\TT{{\cal T}}

\begin{document}
\title{Pattern recognition on random trees associated to functionality families of proteins}
\author{Ana Georgina Flesia }
\address{Ana Georgina Flesia\\FaMAF-UNC\\Ing. Medina Allende s/n, Ciudad Universitaria\\ CP 5000, C\'ordoba, Argentina. }
\email{flesia@mate.uncor.edu}
\urladdr{http://www.famaf.unc.edu.ar/~flesia}
\thanks{AGF was partially supported by PICT 2005-31659. AGF is corresponding author.}
\author{Ricardo Fraiman}
\address{Ricardo Fraiman\\Universidad de San Andrés, Argentina y Universidad de la República, Uruguay }
\email{rfraiman@udesa.edu.ar}
\author{Florencia G. Leonardi}
\address{Florencia G. Leonardi\\Instituto de Matem\'atica e Estat\'istica\\ Universidade de S\~ao Paulo. \\S\~ao Paulo, Brazil.}
\email{leonardi@ime.usp.br}
\thanks{ FGL is supported by FAPESP (grant 06/56980-0). We would like to thank Pablo Ferrari and Antonio Galves for many interesting discussions about metrics in the space of trees and modelling protein functionality.}


\begin{abstract}
In this paper, we address the problem of identifying protein
functionality using the information contained in its aminoacid
sequence. We propose a method to define sequence similarity
relationships that can be used as input for classification and
clustering via well known metric based statistical methods. To
obtain our measure of sequence similarity, we first fit a Variable
Length Markov model to each sequence of our database, generating
estimated context trees, and then we compute the BFFS distance in
tree space between each pair of trees. The BFFS distance takes
into account the structure of each tree, that is directly related
to the most relevant motifs of the sequence, and indirectly, to
the probability of occurrence of each motif. This approach is
motivated by the idea that proteins having the same functionality
could be modelled with the same VLMC, so their estimated context
trees are observations of the same random variable, and should be
close together in tree space. In our examples, we specifically
address two problems of supervised and unsupervised learning in
structural genomics via simple metric based techniques on the
space of trees
 \begin{enumerate}
 \item
 Unsupervised detection of functionality
families via K means clustering in the space of trees, \item
Classification of new proteins into known families via k nearest
neighbor trees. \end{enumerate}

We found evidence that the similarity measure induced by our
approach concentrates information for discrimination.
Classification has the same high performance than others VLMC
approaches.  Clustering is a harder task, though, but our approach
for clustering is alignment free and automatic, and may lead to
many interesting variations by choosing other clustering or
classification procedures that are based on pre-computed
similarity information, as the ones that performs clustering using
flow simulation, see (Yona et al 2000 , Enright et al, 2003) .
\end{abstract} \maketitle

\section{Introduction}
\label{s1}

 A central problem in functional genomics is to determine the function of a protein using only the information contained in its aminoacid chain, Karp (2002).  It is well known that a protein functionality family is formed by proteins that perform the same function on different organisms and by proteins that come from the same organism that have been derived by genetic duplication or rearrangements, Dayhoff (1976), Hegyi (1999). Well characterized proteins within a family may help enhance the process of classification of family members whose functions are not well known or not well understood, Eisenberg (2000). Also, the features characterizing each functionality family may give information about common evolutionary history, Sasson et al (2003).

Most used methods for proposing hypothesis over protein
functionality are based on sequence alignment, Smith (1981). Exact
sequence alignment has a quadratic computational complexity, which
make them unfeasible for large databases. Heuristic methods like
BLAST, (Altschul et al 1997) or FASTA, (Pearson et al 2000)  are
between the most common choices for comparing sequences in large
data sets. Recently, this problem has been addressed also with non
alignment methods, that look for family models with parameters or
characteristics that determine its functionality. An example of
such body of work is the fitting with different markovian models,
like Hidden Markov Models (Rabiner et al 1986), Hidden Markov
Transducers, and Variable Length Markov Chains (Bejerano 2001,
Apostolico et al,  2000). Hidden Markov Models are very powerful
tools for this task, but have the disadvantage of having too many
parameters to fit, and even though,  in practice, they do not
guarantee an optimal choice of model. Recently, Bejerano et al
(2001) proposed to apply Variable Length Markov Chains to the
problem of classification of proteins into families. Some
advantages of this model are the following: it does not depend of
alignment, it has not as many parameters as HMM, and there are
algorithms that can fit the model in linear time, (Apostolico et
al, 2000).

We are going to address two specific problems here:
\begin{enumerate}
 \item Detection of functionality families via sequence clustering,
 \item Classification of new proteins into known families
\end{enumerate}
These problems are directly related to the problem of detecting
protein functionality, so they can be addressed by the same
methods, but with great variations in performance.

From the mathematical point of view, clustering is an ill posed
problem. The definition of functionality family is quite
ambiguous, so it is very difficult to quantify it mathematically
to obtain a unique objective function to optimize. As a result,
computational clustering approaches differ in the representations
of the proteins to be clustered, the definition of the
optimization goals and also in the resulting partitions of the
known protein space. Stability and heterogeneity of the resulting
clusters are known problems that are shared for most methods, but
still they help to build a big picture of the on going
experimental structure which represents super-families. The goal
of fully automated clustering methods becomes to give partial
answers with respect to global organization of all protein
sequences.
  Sequence classification into families is a simpler task than unsupervised learning, but still has delicate problems as the one introduced by multi-domain proteins, and the accuracy of the labelling of the training set. A related problem is to classify sequences belonging to a family into subfamilies that are in most cases, defined by their evolutionary history. Most tailored methods rely on a multiple alignment of the family sequences as well as the phylogeny tree inferred from it. Indeed, when the resulting evolutionary tree can be reconstructed accurately, functional subtypes can often be identified with subtrees within it.

There are many clustering techniques that rely in pairwise
similarity measures of protein sequences. ProtoMap, (Yona et al
2000)  ProtoNet, (Sasson et al 2003), BioSpace (Yona and Levitt
2000), use a combination of the three most common measures of
pairwise similarity, (Smith-Waterman, Fasta and Blast) followed by
the construction of a weighted graph that has the resulting
clusters as the most strong connected components. The evolution of
the graph differs in each algorithm. Tribe-MCL (Enright et
al,2003) uses also BLAST to build up a dissimilarity matrix,
converting it into a probability matrix which is used to simulate
a flow that leads to the final graph. Each algorithm has evolved
into complicated learning machines, to avoid the multi domain
protein problem, and to generate a hierarchical view of protein
space. In this paper we will work with a very simple clustering
machine that also relies in pairwaise similarity, the K means
algorithm, which is well suited to detect generative clusters when
the underlying distributions are concentrated.

In the case of the non alignment methods, most of them rely in
modelling protein functionality, so the more accurate the fit of
the model, the better the results in clustering and
classification. In this paper, we explore a hybrid approach for
protein classification and clustering. We fit a Variable Length
Markov Model to each protein sequence, and we use the architecture
of their associated context trees to perform classification and
clustering, considering a metric on the space of trees.  The BFFS
distance (Balding et al, 2007) takes into account the structure of
each tree, that is directly related to the most relevant motifs of
the sequence, and indirectly, to the transition probabilities
associated to these motifs. This approach is motivated by the idea
that proteins that have the same functionality could be modelled
with the same VLMC. In consequence, their estimated context trees
are observations of the same random element, and should be close
together in tree space. We are combining a model based technique
with the classical similarity based statistical learning.

\section{Material and Methods}

\paragraph{Data handling}
A FASTA file containing all sequences that are to be clustered o
classified into families is assembled. The labels are only visible
for evaluation and training purposes.  This file is transformed
via PST algorithm (Bejerano et al, 2001) in trees of fixed depth
4.
\paragraph{Algorithm}

The file containing all trees is compared against itself using the
BFFS distance for trees. The all against all sequence similarities
generated by this analysis are stored in a square matrix. The
labels are reserved for evaluation and training purposes.

\begin{enumerate}
\item Unsupervised Clustering We have chosen a very simple
clustering technique that rely on distances, and that optimizes
the within sum of distances in each cluster. The distance between
two sequences is in fact the distance between two trees estimated
via the PST algorithm, so we can obtain the partition  C with an
alternating optimization procedure that first compute the cluster
mean centroid trees of a given partition and then reassign the
observations to the closest centroid tree, until the objective
function is no longer decreased. This can be done because there is
a notion of average tree or mean centroid tree that shares
properties of the average in Euclidean space. We have adapted a
Matlab code for K means in order to handle trees in BFS format,
and computing the BFFS distance and the mean centroid tree as
needed.

 \item Supervised Learning
We have approached classification also with one of the simple
schemes that rely on distance. Given a new sequence, we classify
it in the family that has more members between the k closest
members of the database, with a standard code in MATLAB.
\end{enumerate}
\paragraph{Availability}

The original PST algorithm for Variable Length Markov Chain
modelling can be adapted from (Bejerano, 2003). The additional
modules for computing distances and mean trees, necessaries for
protein sequence clustering and classification can be obtained
from the authors upon request.

\section{Variable Length Markov Chain Modeling of protein functionality}

The starting point of our approach to supervised and unsupervised
learning is to learn the structure of the Variable Length Markov
Chain that models each family. We claim in this paper that the
estimator of the context tree that characterize the VLMC of each
family is a random tree, a random object that produces trees
following a distribution that also characterize the family.
Classification and clustering is then carried out in a metric
space of trees, and as it is well known, the success of it depends
strongly in the concentration of the distribution of the family in
tree space. We compute estimates of the VLMC context tree of each
family using the Probabilistic Suffix Trees algorithm.

\paragraph{VLMC and the BFFS space of trees}

A Variable Length Markov Chain is a stochastic process introduced
by Rissanen (1983) in information theory; see also Bühlmann and
Wyner (1999). In this model the probability of occurrence of each
symbol at a given time depends on a finite number of precedent
symbols.  The number of relevant precedent symbols may be variable
and depends on each specific sub-sequence.   More precisely, a
VLMC is a stochastic process $(X_n)_{n\in\mathbb Z}$, with values
on a finite alphabet $\cal A$, such that
  \begin{equation}\label{eq:prob}
    P[X_n=\cdot \,|\, X_{-\infty}^{n-1}=x_{-\infty}^{n-1}] = P[X_n=\cdot
    \,|\,X_{n-k}^{n-1}=x_{n-k}^{n-1}]\,,
\end{equation}
where $x_{s}^{r}$ represents the sequence $x_s,x_{s+1},\dotsc,x_r$ and $k$ is a
stopping time that depends on the sequence $x_{n-k},\ldots,x_{n-1}$.  As the
process is homogeneous the relevant past sequences $(x_{n-k},\ldots,x_{n-1})$ do
not depend on $n$ and are denoted by $(x_{-k},\ldots,x_{-1})$.  Each relevant
past $(x_{-k},\ldots,x_{-1})$ is called a \emph{context}. The set of contexts $\tau$
can be represented as a rooted tree $t$, where each complete path from the
leaves to the root in $t$ represents a context. Calling $p$ the transition
probabilities associated to each context in $\tau$ given by (\ref{eq:prob}), the
pair $(\tau,p)$, called \emph{probabilistic context tree}, has all information
relevant to the model, see Rissanen (1983) and B\"uhlmann et al (1999).

\setlength{\unitlength}{1mm}
\begin{figure}[h]
  \begin{center}
    \begin{minipage}{3cm}
       \begin{picture}(50,58)(-2,-7)
        \put(-2,47){(a)}
        \put(1,5){\includegraphics*[height=3.5cm]{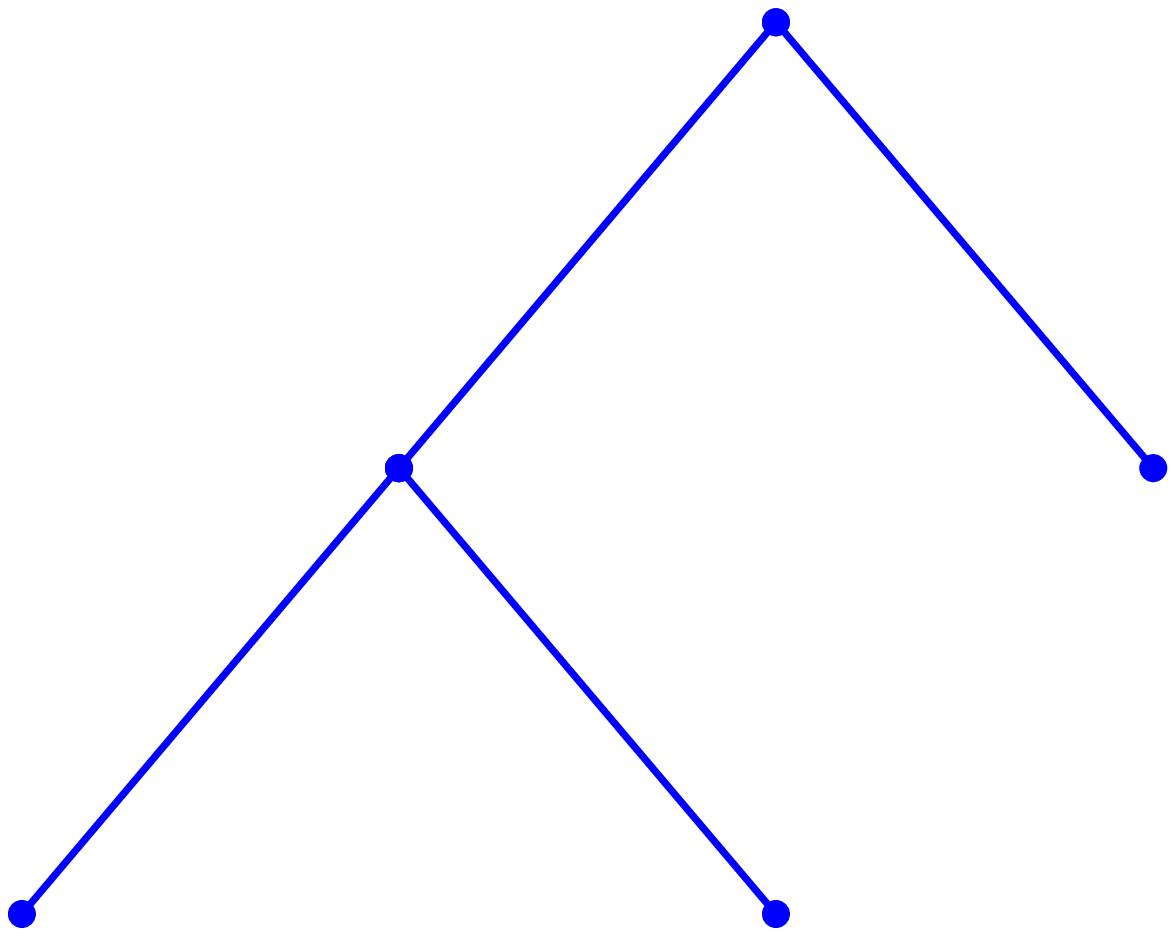}}
        \put(29.5,38.5){\small $\lambda$}
        \put(44,22){\small$\mathbf{2}$}
        \put(44,16){\scriptsize $(0.2,0.8)$}
        \put(15,22){\small 1}
        \put(33,5){\small$\mathbf{21}$}
        \put(33,0){\scriptsize $(0.4,0.6)$}
        \put(3,5){\small$\mathbf{11}$}
        \put(2,0){\scriptsize $(0.7,0.3)$}
      \end{picture}
    \end{minipage}
    \hspace{4cm}
    \begin{minipage}{3cm}
      \begin{picture}(50,58)(-2,-7)
      \put(-2,47){(b)}
        \put(1,5){\includegraphics*[height=3.5cm]{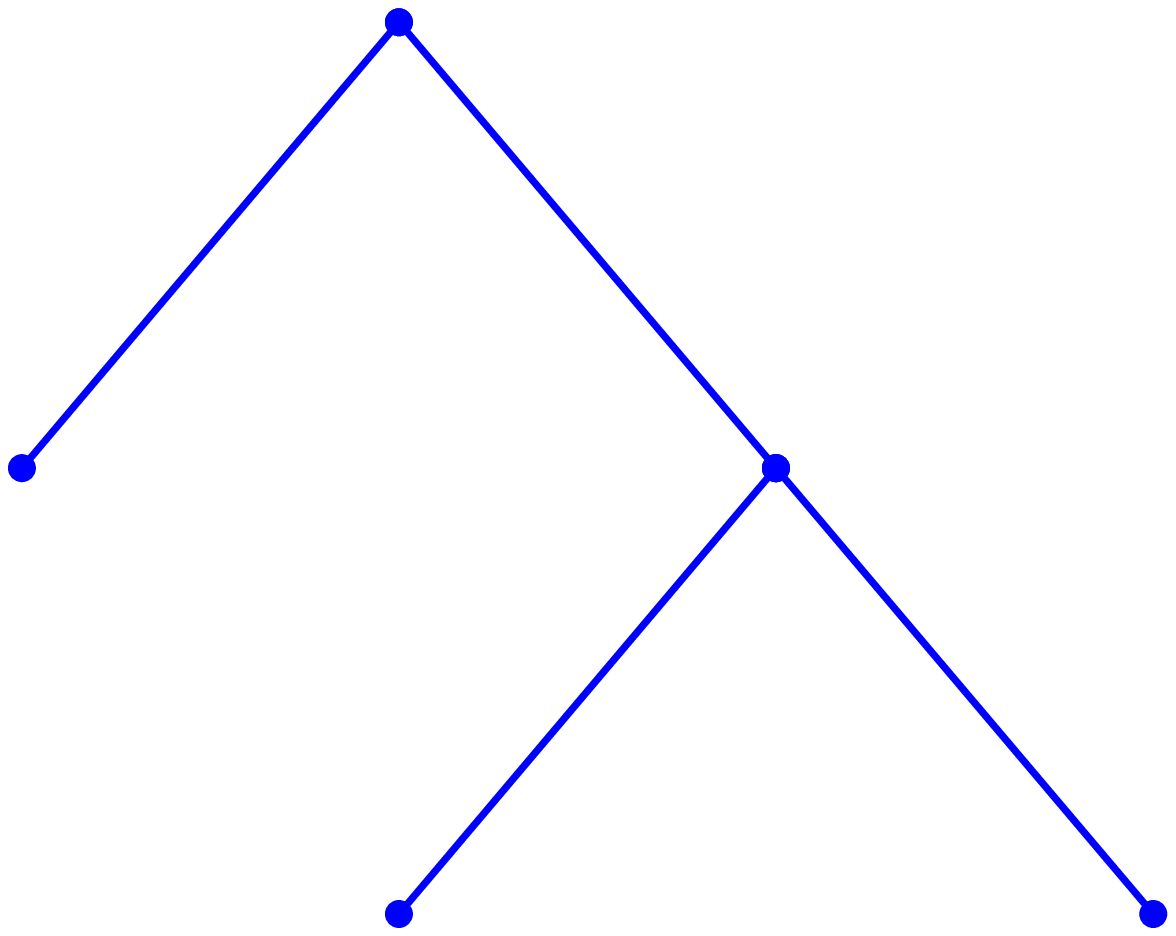}}
        \put(17.5,38.5){\small $\lambda$}
        \put(34,22){\small 2}
        \put(3,18){\scriptsize $(0.6,0.4)$}
        \put(3,22){\small $\mathbf{1}$}
        \put(13,6){\small$\mathbf{12}$}
        \put(45,0){\scriptsize $(0.4,0.6)$}
        \put(44,6){\small$\mathbf{22}$}
        \put(10,0){\scriptsize $(0.2,0.8)$}
      \end{picture}
    \end{minipage}
     \caption{An example of two probabilistic context trees over the alphabet
       $A=\{1,2\}$. (a) The tree $t$
       represents the pair $(\tau,p)$, where $\tau = \{11,21,2\}$ is the set of contexts and $p$ are
       the transition probabilities given by (\ref{e27}).(b) The tree $y$
       represents the pair $(\eta,q)$, where $\eta = \{12,22,1\}$ is the set of contexts and $q$ are
       the transition probabilities given by (\ref{e28}). }
     \label{fig:vlmc}
     \end{center}
   \end{figure}

As an example, take a binary alphabet $\cal A = \{1,2\}$ and transition probabilities
   \begin{equation}
\label{e27}
  P[X_n=x_n \,|\, X_{-\infty}^{n-1}=x_{-\infty}^{n-1}] =
  \begin{cases}
    P[X_n=1 \,|\, X_{n-2}^{n-1} = 1\ 1]& = 0.7, \\
    P[X_n=1 \,|\, X_{n-2}^{n-1} = 2\ 1 ]&=
      0.4, \\
    P[X_n=1 \,|\, X_{n-1}=2]&=0.2.
  \end{cases}
\end{equation}
so that, if $x_{n-1}=2$, then the stopping time $k=1$ and $X_n=1$ with probability $0.2$;
otherwise the stopping time is $k=2$ and $X_n=1$ with probability $0.7$ if both $x_{n-1}=x_{n-2}=1$
or with probability $0.4$ if $x_{n-1}=1$ and $x_{n-2}=2$. The set of contexts is
$\tau = \{11,21,2\}$, when the set of all active nodes of the rooted tree $t$ is $\tV_t = \{11,21,2,1\}$, since $1$ is an internal node in the path of the context $11$ and $21$. Another example over the same alphabet is given by the transition probabilities
\begin{equation}
\label{e28}
  P[Y_n=y_n \,|\, Y_{-\infty}^{n-1}=y_{-\infty}^{n-1}] =
  \begin{cases}
    P[Y_n=1 \,|\, Y_{n-1} = 1]& = 0.6, \\
    P[Y_n=1 \,|\, Y_{n-2}^{n-1} = 2\ 2 ]&=
      0.4, \\
    P[Y_n=1 \,|\, Y_{n-2}^{n-1}=1\ 2]&=0.2.
  \end{cases}
\end{equation}
The set of contexts is
$\eta = \{1,12,22\}$, when the set of all active nodes of the rooted tree $y$ is $\tV_y = \{1,12,2,22\}$, since $2$ is an internal node in the path of the context $12$ and $22$.
The corresponding probabilistic context  trees are
represented in Figure~\ref{fig:vlmc}.

We are going now to embed the set of all possible context trees that
 can be constructed given a fixed alphabet $\cal A$, into the  compact metric
  space of rooted trees, $\TT$, defined by Balding et al (2004).
The relationship is very simple, since a rooted tree can be thought as a subset
of the nodes satisfying the condition ``son present implies father present''.
In this kind of space,
the natural sigma algebra $\BB$ is the minimal one containing cylinders, this is,
 the sets of trees
defined by the presence/absence of a finite number of nodes. The natural
topology is the one generated by the cylinders as open sets. We can associate
to this topology
 a family of distances that take values depending on matched presence/absence
 of nodes in the trees, but not only the leaves, but also the internal ones too.

 So let define $\tV$ as the set of all possible sequences $(x_{-k},\ldots,x_{-1})$
 over the alphabet $\cal A$, with all possible stopping time $k$, and define a tree
 $t$ as a function $t:\tV\to \{0,1\}$ such that $t$ only give value one to the set of
  contexts of the tree and the set of internal nodes of these contexts.
  With this definition, it is easy to define our distance as
\begin{eqnarray}
\label{d12}
d(t,y)=\sum_{v\in\tV}\phi(v)(t(v)-y(v))^2
\end{eqnarray}
where $\phi:\widetilde V\to\R^+$ is a strictly positive function such that
$\sum_{v\in \widetilde V} \phi(v)<\infty$.
In particular we use the function
\begin{equation}
  \label{phi1}
  \phi(v) = z^{{\rm gen}(v)}\,,
\end{equation}
with $z=0.1$ and $ {\rm gen}(v)$ stands for the generation of the node, the number of symbols to reach the root. Let compute the distance between the two trees given in the preceding example,
\begin{eqnarray*}
d(t,y)&=&\sum_{v\in\tV}\phi(v)(t(v)-y(v))^2\\&=&\phi(\lambda)(t(\lambda)-y(\lambda))^2
+\phi(1)(t(1)-y(1))^2+\phi(2)(t(2)-y(2))^2+\phi(11)(t(11)-y(11))^2\\&&
+\phi(12)(t(12)-y(12))^2+\phi(21)(t(21)-y(21))^2+\phi(22)(t(22)-y(22))^2\\
&=&0+0+0+\phi(11)
+\phi(12)+\phi(21)+\phi(22)\\
&=&4\times 0.1^3=0.004
\end{eqnarray*}

With this distance $(\TT,d)$ becomes a compact metric space, see details in Balding et al(2004). The computation of the distance is not as fast a s the computation of the Euclidean distance but it may be devised a code that search for coincidences by contexts, that means, going from root to leaves.

\paragraph*{\bf Random trees}
A \emph{random tree} with distribution $\nu $ is a measurable
function
\begin{equation}
  \label{p3}
  T:\Omega\to\TT\;\hbox{such that}\; \P(T\in A) = \int_A \nu(dt) \;.
\end{equation}
for any Borel set  $A\in \BB$, where $(\Omega,\mathcal F,\P)$ is a
probability space and $\nu$ a probability on $(\TT,\BB)$.

Given a sample of independent random trees $T_1,\dots,T_n$ with identical distribution $\nu$ on our compact metric space $(\TT,d)$, a measure of central tendency is a {\em sample centroid} defined as a tree (or set of trees) ${\bf \overline T_n}$ in $\TT$ satisfying

 \begin{equation}
{\bf \overline T_n }:= \arg\min_{t\in\TT} \frac1n\sum_{i=1}^nd(T_i,t).
 \end{equation}

This formula may show the problem as more difficult that it is,
since it is calling for a search over the whole set of trees, that
grows exponentially in the number of nodes. But it is easy to
prove that the sample centroid (or mean tree) of a set of trees
can be built by majority vote over the nodes. That means, at least
one of the sample centroids (it does not need to be unique) can be
defined as the tree whose nodes are present only if they are
present in at least half of the sample.

 \section{Supervised and
unsupervised learning on the space of trees }

In cluster analysis, the goal is to find an optimal partition for
which observations or objects within each cluster are similar, but
the clusters are dissimilar to each other. It differs
fundamentally from classification analysis, where the observations
are allocated to a known number of predefined groups or
populations. Many techniques are based on a certain measure of
similarity between pairs of observations.

\paragraph{K means clustering}
We are concerned with a particular cluster technique, called K
means clustering procedure, which generates the class labels
trough the minimization of the "within cluster" point scatter, a
dissimilarity based loss function defined by
\[
W(C)=\frac{1}{2}\sum_{k=1}^K\sum_{C(i)=k}\sum_{C(i')=k}\|x_i-x_{i'}\|^2
\]
This rule characterizes the extent to which observations assigned
to the same cluster tend to be close to each other. It was
initially intended for real valued quantitative variables, and the
squared Euclidean distance was chosen as a measure of
dissimilarity.   In our case, we have objects belonging to a space
of trees, so we choose the distance d as the dissimilarity
measure, and the within-point scatter may be redefined as
\[
W(C)=\sum_{k=1}^K\sum_{C(i)=k}d(t_i,\overline{\bf t_k})
\]
 where $\overline{\bf t_k}$ is the sample centroid associated
 to the $k$th cluster. As in the case of the Euclidean space, an
 iterative descent algorithm for solving
 \[
 C^*=\mbox{arg} \min_{C}\sum_{k=1}^K\sum_{C(i)=k}d(t_i,\overline{\bf t_k})
 \]
 may be obtained by noting that for any set of observed trees $S$
 \[
 \overline{\bf t_S}=\mbox{arg} \min_t\sum_{t_i\in S}d(t_i,t)
 \]
 by definition. Hence we can obtain $C^*$ with
 an alternating optimization procedure that first compute
 the cluster mean centroid trees and then reassign the observations
 to the closest centroid tree, until the objective function is no longer decreased.
  This one of the most popular iterative descent algorithms that go by the name of
  K means, and the one we use in our examples.  It is not difficult to prove, following
  Pollard(1981), par example, that in the case of locally compact metric
  spaces the sample centroid trees converge to the population centroid tree when the sample
   increases. This a great result that ensures that if there is a partition of
   the population into $K$ clusters, and  if we have enough data, with the proper
   initialization to avoid local minima, the K-means algorithm will give an
    accurate outline of the clusters.

As a final note we state that $K$ means algorithm could be used for classification purposes, following the next steps
\begin{itemize}
\item apply $K$ means clustering to the training data in each of the $K$ classes separately, using $R$ prototypes per class
\item assign a class label to each of the $K\times R$ prototypes
\item classify  a new feature $t$ to the class of the closest prototype.
\end{itemize}

    This is an example of the prototype methods of classification that
    can be also adapted to work on spaces of trees. The difference between
     this method and k-Nearest Neighbors is the fact that the prototypes are
     not part of the training samples, but the centers of the partition of each
      training sample class.

\paragraph{k-nearest neighbors classification}

Given a family of proteins $\cal F$ and a new sequence of amino
acids $s$, the goal is to determine if s belongs to $\cal F$ or
not. To answer that question, Bejerano (2003) and Leonardi (2007)
estimate first a model for the family $\cal F$, using sequences
classified inside the family. To determine the label of a new
protein, they search for the family whose model has higher
probability of having produced that sequence. The model
constructed for the family $\cal F$ is a Variable Length Markov
Chain, obtained estimating the probabilistic tree that matches the
chain by means of the PST algorithm as in Bejerano (2003). In this
paper we fit the simplest model for classification, we consider
that proteins from the same family are clustered tightly,
measuring it with the BFFS distance for context trees, and we
score the new protein with the rule of the $k$-nearest neighbours.
We label the protein as belonging to the family that has more
neighbours in this $k$-subjects neighbourhood. The $k$-nearest
neighbour rule is a very simple, distance based method for pattern
recognition or data classification. This method relies on the
intuitive concept that data points of the same class should have
neighbours in the class (in distance) with high probability.  As a
result, for a given data point of an unknown class, we can simply
compute the distance of this point to the training data, and
assign the class determined using majority vote among the $k$
neighbours of this data point. The algorithm is straightforward:

\begin{itemize}
\item Given a new observation $t$ we find the $k$ training points $S=\{t_{(1)},\dots,t_{(k)}\}$ closest in distance to $t$.
\item Classification is made using majority vote among the $k$ neighbors in $S$.
\item
 Ties are broken at random.
\end{itemize}

The simplicity of the rule is extreme, and its success depend only
in the ability of the measure to cluster families, and the ability
of the VLMC modelling to successfully detect all the differences
between the protein chains of the same family and resume them into
its context tree.

\section{Computational Examples}
In this section we would like to assess the capability of VLMC
methods to capture the essential structure of the family that
would help the discrimination problem. Traditional PST
classification methods choose a training set of sequences of a
given family and estimates the context tree of the family
concatenating all the sequences in that training set. Then
classification is performed computing the probability that a given
sequence would be produced by that context tree. The motivation of
such approach is related to biological understanding of the
evolution and composition of protein families. We suppose that a
group of evolutionary related protein sequences should exhibit
many identical short segments which have been either preserved by
selection or have not diverged long enough from their common
single ancestral sequence. The variable memory model is well
equipped to pick up these locally conserved segments, showing them
in the architecture of the context tree, (Bejerano, 2003). Our
approach diverges from the classical approach of VLMC methods
since we do not use for classification the empirical probabilities
associated to each context but the architecture of the context
tree that is computed for each protein sequence in the family.
Also, we do not collect all the samples to generate an estimation
of the model, but we compute an estimate per sample sequence and
look how they cluster together in tree space. The context tree
built with all the collected sequences will show segments that are
consistently repeated in most of the sequences, but context trees
built with each sequence will show patterns inside each particular
sequence, and the family bond will emerge as a relationship in
tree space. The definition of the distance is thus fundamental for
our approach. We have illustrated these ideas with some small
examples, following the methodology of Bejerano (2003), since we
want to compare our approach with the traditionally PST approach,
and indirectly, with other classification methods. Our reference
is the Pfam database that is based on Hidden Markov Models trained
in a smaller database of manually curated well known proteins.

\paragraph{K means}

\begin{figure}[h]
\centering
\includegraphics*[height=7cm]{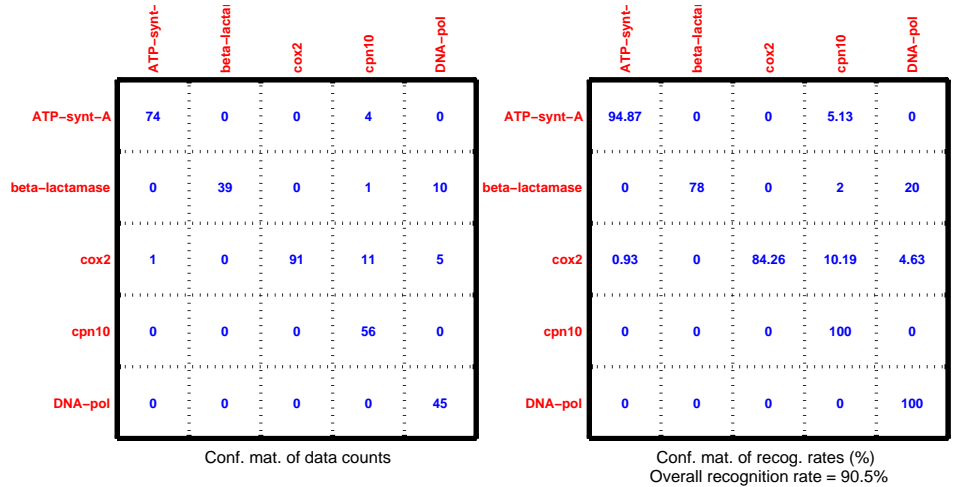}
\label{figmat1} \caption{Confusion matrices of data counts and
percentage of well classified and misplaced proteins over 5
families of proteins}
      \end{figure}

\begin{figure}[h]
\centering
\includegraphics*[height=7cm]{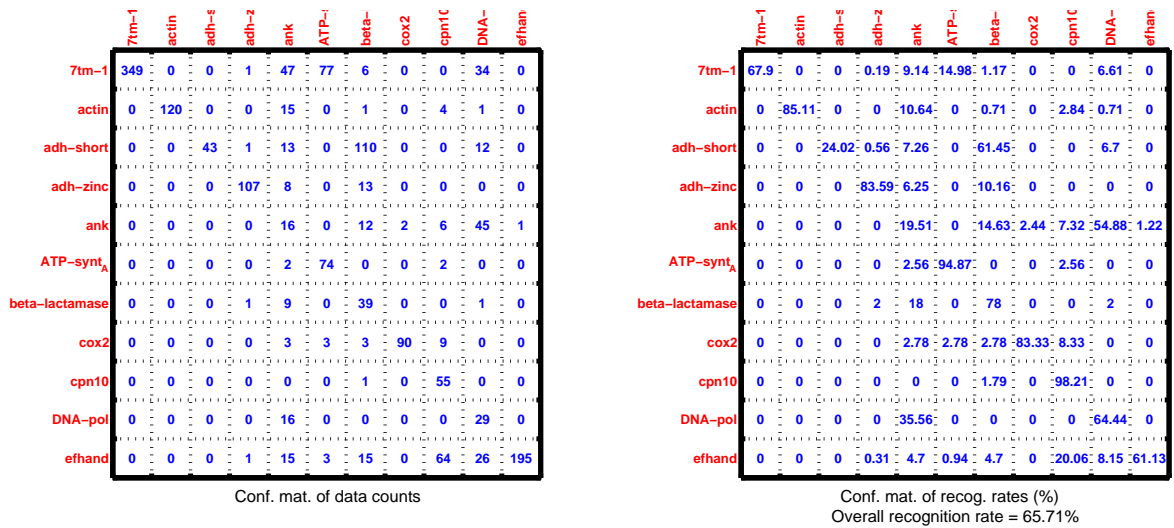}
\label{figmat2} \caption{Confusion matrices of data counts and
percentage of well classified and misplaced proteins over 11
families of proteins }
      \end{figure}

Clustering is a very hard problem, since there is no help of
ground truth to shape the families. Most databases have been
constructed clustering sequences with statistical methods, but
with the help of a core of manually curated well known proteins.
We do not aim to show a method that would accurately cluster the
whole Pfam database in seconds, but to discuss the potential that
our model could have for checking coherence and relationships
among families. We should notice too that K means is an algorithm
that considers similarities in a pairwise fashion, methods that
consider global similarities could help cluster assignment.
In order to assess the ability of the K means procedure to
 cluster proteins in tree space using the BFFS distance, we have
  started transforming  5 families of proteins selected from the Pfam
   database, labelled  'ATP-synt-A', 'beta-lactamase','cox2', 'cpn10', 'DNA-pol'.
   The overall performance is 90.5 \%. In Figure \ref{figmat1} we have plot two matrices,
    one of counts and other with percentage of well classified and misplaced proteins.
We notice that the two families that are correctly clustered,
showing a high degree of coherence, also attract members from
other families, as it is shown in Figure \ref{figmat1}. In our
second example we added 6 more families,
'7tm-1','actin','adh-short','adh-zinc','ank', and 'efhand',
 and the overall recognition rate drop to 64 \%,
 but the confusion matrix plot in Figure \ref{figmat2} show us that proteins are
 not scattered around but are misplaced in specific families. This feature
 could be very interesting at the time to determine the coherence of the family
 and the relationships between different families. For example, from the
 ten proteins of the beta lactamase family that have been incorrectly assigned to
 the 'Dna-pol' family, 9 of them have been reassigned now to the 'ank' family, but
  50\% of the ank proteins have been assigned to the 'Dna-pol' family, showing
   that these three families are  close in tree space.

\paragraph{k nearest neighbors}

\begin{figure}[h]
\centering
\includegraphics*[height=5cm,width=12cm]{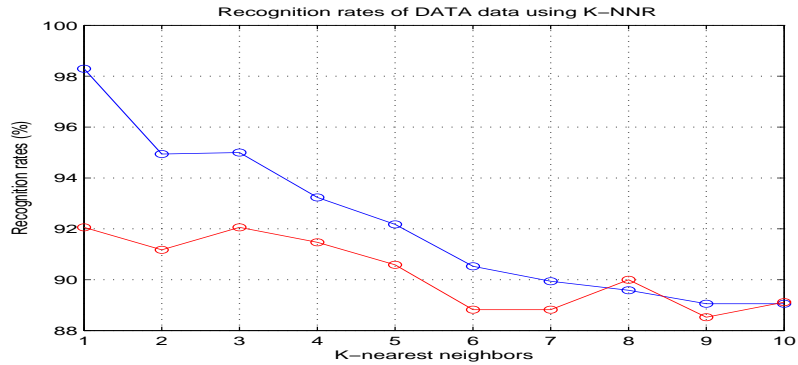}
\label{knn11} \caption{In blue, we plot the overall recognition
rates of $k$ nearest neighbors rule as a function of k, considering a
 training set of 80\% of the total of proteins, classifying the 100\%.
  In red, we plot the overall
recognition rates computed classifying only the 20\% set aside  from the training set.}
      \end{figure}
Now we consider the same 11 families from the previous example, '7tm-1','actin',
 'adh-short','adh-zinc','ank','ATP-synt-A','beta lactamase', 'cox2','cpn10', 'DNA-pol',
 'efhand' but we easy the difficulty of the clustering problem allowing the help of a
  training set. We selected randomly 80\% of the whole set of proteins,
   and set it aside as training set. We could refine later the example,
    choosing randomly 80\% of each family as a training set.
      Then we classify the whole set of 1700 proteins using the training set
      to generate neighborhoods of different sizes, and plot in blue the percentage
       of true positives detected as a function of the number of neighbors considered.
        Secondly, we compute the percentage of true positives detected but considering
         only the 20\% of proteins that are not included in the training set,
         and plot it in red as a function of the number of neighbors considered,
         see Figure \ref{knn11}. The last rate computed is a better indicator of
          the performance of the rule, since any rule should do well in the train set.
           In fact, usually, one nearest neighbor achieves 100\% detection in the train set.
           Even though, we want to compare with the PST classification rule itself,
           and the first rate is the one reported in Bejerano (2003).

Our report will start stating that the 98\% of the whole set
(counting the training set) has been well classified using the one
nearest neighbour rule, against a 60\% of true positives given by
K means. If we do not count the training set, the 92\% of the test
set is well classified with one nearest neighbour rule. It is
interesting to notice that if we allow more than three points in
the neighbourhood, we have less accuracy in the classification,
giving evidence to the idea that the differences between the trees
are very subtle. We have written in Table \ref{Tabla4} the
percentage of true positives per family as reported in Bejerano
(2003). We should point out that we are using very short trees,
and still we achieve the same rates of classification. The first
three columns KNN1, KNN2 y KNN3 are the percentage of good
classification training with 1, 2 and 3 neighbours, computed using
a training set of 80\% random proteins and scoring 100\%.  The
last three columns show the same variables, but computed only over
the 20\% of samples that are not part of the training set. The
percentages are reduced in some cases, but not very much, but the
interpretability of the results and credibility of the experiment
has been reinforced. We only compute these first three columns to
compare with PST values, which are computed in this fashion.
\begin{table}[h]
\small
\begin{center}
  \begin{tabular}{|c|c|c|c|c|c|c|c|c|c|c|}
      \hline
      \hline
      Family &Size&Coverage&\%PST  &\%KNN1 &   \%KNN2&   \% KNN3 &\%KNN1 &   \%KNN2&   \% KNN3\\
      \hline
  \verb+7tm_1+ & 515         & 0.707&93.0 &99.8   &99.4   &98.8&   99.0 &  99.0 & 100.0   \\
  \verb+7tm_2+ & 36          & 0.735&94.4 &100    &100    &94.2&  100  &  100.0&  100.0\\
  \verb+7tm_3+ & 12          & 0.805&83.3 &100    &100    &100 &  100  &  100.0&  100.0\\
  \verb+AAA+ & 66            & 0.378&87.9 &98.4   &90.7   &89.2&   92.3&  100.0&  100.0\\
  \verb+ABC_tran+ &269       & 0.518&83.6 &95.5   &87.6   &88.4&   77.7&   74.0&   74.0\\
  \verb+actin+ & 142         & 0.965&97.2 &100    &98.5   &98.5&  100.0&  100.0&  100.0\\
  \verb+adh_short+ & 180     & 0.661&88.9 &95.5   &87.1   &89.9&   77.7&   75.0&   80.5\\
  \verb+adh_zinc+ &129       & 0.970&95.3 &97.6   &90.6   &91.4&   92.3&   92.3&   88.4\\
  \verb+aldedh+ & 69         & 0.907&87.0 &98.5   &95.5   &92.6&   92.8&   85.7&   92.8\\
  \verb+alpha-amylase+ &114  & 0.750&87.7 &98.2   &92.9   &92.0&   91.3&   78.2&   86.9\\
  \verb+aminotran+ &63       & 0.942&88.9 &95.1   &74.1   &77.4&   76.9&   46.1&   69.2\\
  \verb+ank+ &83             & 0.151&88.0 &92.6   &74.3   &63.4&   64.7&   58.8&   47.0\\
  \verb+arf+ &43             & 0.951&90.7 &100    &97.6   &100 &  100.0&  100.0&  100.0\\
  \verb+asp+ & 72            & 0.771&83.3 &97.1   &97.1   &94.3&   86.6&   93.3&   86.6\\
  \verb+ATP-synt_A+ &79      & 0.649&92.4 &100    &96.1   &96.1&  100.0&  100.0&  100.0\\
  \verb+ATP-synt_ab+ &180    & 0.694&96.7 &100    &97.7   &99.4&  100.0&  100.0&  100.0\\
  \verb+ATP-synt_C+ &62      & 0.855&91.9 &100    &96.7   &98.3&  100.0&  100.0&  100.0\\
  \verb+beta-lactamase+ &51  & 0.863&86.3 &98     &92     &90  &   90.0&   90.0&   90.0\\
  \verb+bZIP+ & 95           & 0.217&89.5 &96.8   &86.1   &85.1&   84.2&   78.9&   78.9\\
  \verb+C2+ & 78             & 0.175&92.3 &94.8   &66.2   &83.1&   75.0&   75.0&   68.7\\
  \verb+cadherin+ &31        & 0.503&87.1 &100    &93.3   &100 &  100.0&  100.0&  100.0\\
  \verb+cellulase+ &40       & 0.584&85.0 &92.3   &76.9   &76.9&   62.5&   50.0&   62.5\\
  \verb+cNMP_binding+ &42    & 0.466&92.9 &100    &90.2   &87.8&  100.0&  100.0&  100.0\\
  \verb+COesterase+ & 60     & 0.900&91.7 &98.3   &93.2   &93.2&   91.6&   83.3&   75.0\\
  \verb+connexin+ & 40       & 0.687&97.5 &100    &100    &100 &  100.0&  100.0&  100.0\\
  \verb+copper-bind+ & 61    & 0.835&95.1 &100    &98.3   &100 &  100.0&  100.0&  100.0\\
  \verb+COX1+ & 80           & 0.214&83.8 &100    &100    &96.2&  100.0&  100.0&  100.0\\
  \verb+COX2+ &109           & 0.897&98.2 &99.0   &97.2   &96.3&   95.4&   90.9&   90.9\\
  \verb+cpn10+ &57           & 0.953&93.0 &100    &98.2   &98.2&  100.0&  100.0&  100.0\\
  \verb+cpn60+ & 84          & 0.948&94.0 &100    &100    &100 &  100.0&  100.0&  100.0\\
  \verb+crystall+ &53        & 0.851&98.1 &100    &100    &100 &  100.0&  100.0&  100.0\\
  \verb+cyclin+ & 80         & 0.635&88.8 &94.9   &89.8   &87.3&   75.0&   75.0&   68.7\\
  \verb+Cys_knot+ &61        & 0.502&93.4 &96.6   &98.3   &95  &   83.3&   83.3&   91.6\\
  \verb+Cys-protease+ &91    & 0.682&87.9 &95.5   &94.4   &91.1&   77.7&   77.7&   77.7\\
  \verb+cystatin+ &53        & 0.742&92.5 &98.0   &90.3   &82.7&   90.9&   81.8&   72.7\\
  \verb+cytochrome_b_C+ &130 & 0.313&79.2 &85.2   &65.8   &74.4&   26.9&   50.0&   46.1\\
  \verb+cytochrome_b_N+ &170 & 0.658&98.2 &45.5   &62.7   &53.8&   26.4&   32.3&   29.4\\
  \verb+cytochrome_c+ &175   & 0.891&93.7 &96.5   &95.4   &95.4&   88.5&   85.7&   88.5\\
  \verb+DAG_PE-bind+ &68     & 0.112&89.7 &58.2   &70.1   &62.6&   42.8&   57.1&   50.0\\
  \verb+DNA_methylase+ & 48  & 0.846&83.3 &93.6   &85.1   &87.2&   70.0&   60.0&   80.0\\
  \verb+DNA_pol+ & 46        & 0.650&80.4 &97.7   &91.1   &91.1&   88.8&   77.7&   77.7\\
  \verb+dsrm+ & 14           & 0.226&85.7 &92.3   &69.2   &76.9&   66.6&   33.3&   33.3\\
  \verb+E1-E2_ATPase+ & 102  & 0.636&93.1 &96.0   &90.0   &87.1&   80.9&   76.1&   76.1\\
  \verb+efhand+ & 320        & 0.401&92.2 &95.9   &94.6   &92.7&   92.1&   92.1&   92.1\\
  \verb+EGF+ & 169           & 0.133&89.3 &92.8   &84.5   &85.7&   85.2&   85.2&   70.5\\
  \verb+enolase+ & 40        & 0.983&100  &97.4   &97.4   &97.4&   87.5&   87.5&   87.5\\
  \verb+fer2+ & 88           & 0.785&94.5 &97.7   &93.1   &90.8&   88.8&   88.8&   94.4\\
  \verb+fer4+ & 152          & 0.559&88.2 &97.3   &90.7   &90.7&   87.0&   70.9&   83.8\\
  \verb+fer4_NifH+ &49       & 0.928&95.9 &100    &97.9   &97.9&  100&  100&  100\\
  \verb+FGF+ & 39            & 0.691&97.4 &100    &97.3   &100  &  100&  100&  100\\
\hline\hline
\end{tabular}
\end{center}
\label{Tabla4} \caption{Family name, size: number of proteins in
it, percentage of correct classification of
 PST method, percentage of classification of KNN1, KNN2 and KNN3 using 80\% of the samples as
 a training test, and 100\% as a test set, and 20\% aside from training as test set. }
\end{table}

\section{Final Remarks}

Pattern recognition is an active field of research within
engineering and computer science communities. Its main goal is to
develop automatic methods for recognizing patterns in data. It
encloses sub disciplines like discriminant analysis, feature
extraction, error estimation and cluster analysis, among others.
There are two specific methods that are well known in the
literature of statistical pattern recognition that we have
refereed here. They are \begin{enumerate} \item K means clustering
\item k nearest neighbors classification
\end{enumerate} We have addressed these two methods not
in the conventional setting of a Euclidean space, but when the
databases we have to classify and cluster consist of finite trees.
We have considered a compact metric space of trees as the natural
space where our database lies, and we have shown extensions of the
two aforementioned procedures that apply in this new space. An
interesting example of such database of trees is the one obtained
when a general set of codified strings is modelled with a Variable
Length Markov Chain. The VLMC is represented by its context tree,
which can be estimated from each string using an algorithm like
PST (Bejerano, 2003) or Context (Rissanen, 1983) leading to the
final database of estimated trees. If the codification is correct,
we claim that the context tree of the chain will have all the
information that is needed for discrimination by metric based
methods.
 In functional genomics, proteins are codified as strings of amino acids,
  and VLMC models are naturally fitted to functional families of such strings.
   Amino Acid chains are natural candidates for this type of modelling, but any
   suitable codification of an object with a finite alphabet will make this model arise,
    so other problems besides functional genomics could make profit of this type of approach.
     In classification of written reports, codification of the reports is usually made
     in order to reduce dimensionality or to extract features, leading to a database of
     strings, see Jeske and Liu (2004) and Jeske and Liu (2006). In these papers,
      the codification is derived carefully to ensure discrimination into "bad reports" or
       "good reports".  But also, the codification could represent
        a conjecture made over style or prosody of speech or written text
        as it has been done by  Veilleux et al (1990) in a general setting, and Dorea
         et al (1997), and Frota et al (2001) for the case of detecting differences
          between Brazilian and European Portuguese. In this case, successful discrimination
           may give evidence to support the linguistic conjecture.
            We believe that our context can be of great importance for
             addressing several problems of Computational Linguistics, Natural
             Language Modelling and Speech Processing.

\end{document}